\documentstyle[prl,aps,epsfig,rotating,12pt]{revtex}

\fontsize{12}{16}

\def\bild#1#2{\epsfig{file=#1,width=#2}}

\def\beq{\begin{equation}}
\def\eeq{\end{equation}}

\def\eps_null{\varepsilon_0}

\begin{document}

\title{Harmonic generation in ring-shaped molecules}
\author{F.~Ceccherini and D.~Bauer}
\address{Theoretical Quantum Electronics (TQE), Darmstadt University of Technology,\\ Hochschulstr.\ 4A, D-64289 Darmstadt, Germany}
\date{\today}

\maketitle

\begin{abstract} 

\begin{center} ABSTRACT  \end{center}

\noindent We study numerically the interaction between an intense circularly polarized laser field and an electron moving in a potential which has a discrete cylindrical symmetry with respect to the laser pulse propagation direction. This setup serves as a simple model, e.g., for benzene and  other aromatic compounds. From general symmetry considerations, within a Floquet approach, selection rules for the harmonic generation [O.\,Alon {\em et al.}, Phys. Rev. Lett. {\bf 80}, 3743 (1998)] have been derived recently. Instead, the results we present in this paper have been obtained solving the time-dependent Schr\"odinger equation {\em ab initio} for realistic pulse  shapes. We find a rich structure which is not always dominated by the laser harmonics. 
\end{abstract}
\vspace*{1cm}

\newpage

\section{Introduction}
\noindent The generation of harmonics (HG) through the interaction of atoms with intense laser fields is a topic which has been broadly studied from both a theoretical and an experimental point of view. The big interest in the HG is due to the possible use as a source of short-wavelength radiation. In fact, through the harmonic emission it is possible to generate coherent XUV radiation using table-top lasers. Recently, harmonics of wavelength as short as 67 {\AA} have been reported \cite{preston}. The harmonics spectra obtained from a single atom in a monochromatic laser field present some common and well known features: (i) only linearly polarized odd harmonics are generated \cite{bloem}, (ii) the spectrum has a plateau structure , (iii) the plateau is extended up to a cut-off that is located around $I_p + 3.17 U_p$, where $I_p$ is the ionization energy of the atom and $U_p$ is the ponderomotive energy \cite{krause}. The presence of only odd harmonics is due to symmetry reasons (a more detailed argument will be discussed in the next Section) and the location of the cut-off can be explained, at least qualitatively, with the so-called ``simple man's theory'' \cite{simple}. The interaction of a single atom with two circularly polarized lasers of frequencies $\omega$ and $2\,\omega$ has been investigated recently \cite{becker}; it has been found that while the harmonic orders $3\,n-1$ and $3\,n+1$ are allowed, the harmonic order $3\,n$ is forbidden by the selection rules of the dipole emission, where $n$ is any positive integer. These results agreed with  a previous experiment \cite{eichmann}. 

\noindent More recently, the generation of harmonics in more complex systems than the single atom has become a strongly addressed topic. A model for harmonic emission from atomic clusters has been proposed \cite{maquet}. The harmonics generated by electrons moving in the periodic potential of a crystal have been investigated also \cite{roso,faisal}.   

\noindent In this work we want to study the generation of harmonics in ring-shaped molecules, like benzene and other aromatic compounds. This kind of molecules exhibits an invariance under a rotation of a certain angle around the axis which is orthogonal to the molecule plane and goes through its center. In this case the potential is periodic in the azimuthal direction.

\noindent The HG from ring-like molecules interacting with a circularly polarized field presents many different features with respect to the single-atom case in a linearly polarized field: (i) within the same harmonic range fewer lines are emitted and the higher is the number of atoms in the molecule, the lower is the number of emitted lines, (ii) odd and even harmonics are equally possible, (iii) the  harmonics are alternately left or right circularly polarized. In our opinion, all these peculiar properties make this topic challenging and worth to be studied in detail. 

\noindent The paper is organized as following: in Section II we summarize the derivation of the selection rules for the ring-shaped molecules obtained by Alon {\em et al.} \cite{alon}. In Section III the numerical model used in our simulations is presented and discussed. In Section IV we describe the interaction between the ring molecule and the laser field. In Section V we show the results obtained for different intensities and frequencies together with a broad discussion. Finally, in Section VI we give a summary and an outlook. Atomic units (a.u.) are used throughout the paper.

\section{Selection Rules}

\noindent In the case of atoms or molecules which are shone by a laser
field of frequency $\omega$,  the Hamiltonian is periodic in time
with a period $\tau = 2\pi/\omega$: $H(t+\tau) = H(t)$. 
The time-dependent Schr\"odinger equation (TDSE) for such a system can be written as

\beq
\left [H(t) -i \frac{\partial}{\partial t} \right ] \Psi_{\cal W}(\vec{r}, t) = 0.
\eeq
\noindent The solutions are of the form (Floquet theorem):
\beq
  \Psi_{\cal W} (\vec{r}, t) = \Phi
  (\vec{r}, t)\mbox{e}{}^{-i {\cal W} t} \hspace{0.5cm} \makebox{with} \hspace{0.5cm} \Phi (\vec{r}, t + \tau) =  \Phi (\vec{r}, t), 
\eeq
where ${\cal W}$ is the quasi-energy and $ \Phi (\vec{r}, t)$ is a square integrable
function.\\
Because the set of all functions which are square-integrable in a certain interval and have a finite norm over a cycle, forms a composite Hilbert space, we can apply the extended Hilbert space formalism. The probability to get the $n$th harmonic from a system in a state
$ \Psi_{\cal W} (\vec{r}, t) $ is \cite{sambe}

\beq
\sigma^{(n)}_{\cal W} \propto n^4 \bigg| \langle \langle \Phi (\vec{r}, t) |
\hat{\mu} e^{-i n \omega t} | \Phi (\vec{r}, t) \rangle \rangle  \bigg |^2,
\eeq

\noindent where $\hat{\mu}$ is the dipole  operator and the double
bracket stands for the integration over space and time. 

\noindent In the case of an atom (in the dipole approximation) the Hamiltonian ${\cal H}(t) \equiv [H(t) -i \frac{\partial}{\partial t} ]$ is invariant under the so-called second order dynamical symmetry operator (DSO) \cite{alon},

\beq
 {\cal P}_2 = (\vec{r} \rightarrow -\vec{r}, t
  \rightarrow t + \pi/\omega).
\eeq
\noindent Therefore, the states $| \Phi \rangle \rangle $ are simultaneous
eigenfunctions of ${\cal H}(t)$ and  the second order DSO with
eigenvalues $\pm 1$. The $n$th harmonic is therefore emitted only if 
\beq
\langle \langle \Phi | \hat{\mu}e^{-in\omega t} | \Phi \rangle \rangle =
\langle \langle {\cal P}_2 \Phi | {\cal P}_2 \, \hat{\mu} e^{-i n \omega t
  }\, {\cal P}_2^{-1} | {\cal P}_2 \Phi \rangle \rangle \neq 0
,
\eeq
\noindent leading to 
\beq
\hat{\mu}(\vec{r})
  e^{-i n \omega t} = \hat{\mu}(-\vec{r}) e^{ -i n \omega (t + \pi/\omega)} 
\eeq
\noindent that is fulfilled only with {\em odd} $n$'s. Instead, where the Hamiltonian ${\cal H}(t)$
is invariant under a rotation around an
  $N$-fold symmetry axis ${\cal P}_2$ can be replaced by \cite{alon}
\beq
{\cal P}_N = \left ( \varphi \rightarrow \varphi + \frac{2\pi}{N}, t
  \rightarrow t + \frac{2\pi}{N \omega} \right ),
\eeq
where $\varphi$ is the angular coordinate around the symmetry axis. With an
  algebra similar to the single atom case we derive 
\beq
e^{\pm i \left (\varphi + \frac{2\pi}{N} \right )}e^{-i n \omega \left (t +
    \frac{2\pi}{N \omega} \right )} = e^{\pm i \varphi}e^{-in\omega t}
    \Longrightarrow e^{-i \frac{2\pi \left (n \pm 1 \right )}{N} } = 1,
\eeq
\noindent from which $ n = kN\pm 1, k \in {\cal N}$ follows. That means that the higher is the symmetry order $N$ the less are the
generated harmonics within a fixed frequency interval. In the limit of a continuous symmetry $C_{\infty}$ a
circularly polarized laser does not generate any harmonics. The two harmonics of
each couple have opposite polarization, clockwise and anticlockwise \cite{polar}.

\section{Numerical Model}

\noindent In order to keep the numerical effort manageable we restrict ourselves to a two-dimensional (2D) model where the molecule plane and the rotating electric field are properly represented. We study a single active electron in a ring-shaped potential of $N$ ions. Different kinds of ``smoothed'' potentials can be used for this purpose \cite{smooth}. The potential used in our simulations reads

\beq
V(\rho, \varphi) = -\frac{A}{\sqrt{(\rho-r_0)^2 + \beta}}(\alpha\,\mbox{cos}\,(N\,\varphi)+2-\alpha)
\eeq

\noindent where $r_0$ is the radius of the molecule, and $\rho$ and $\varphi$ are the polar coordinates. $\beta$ is the parameter which gives the ``degree of smoothness'' of the potential and determines the width of the ground state along the ring. $\alpha$ moves the local maxima of the potential keeping the minima constant (the parameter $\alpha$ was introduced in order to avoid the presence of points where $V=0$ for finite $\rho$, because this could generate non-physical ionization). Finally, $A$ is the  ``strength'' of the potential. For our simulations we chose $\alpha = 0.99$ and $\beta = 0.38$. Once the values of $\alpha$ and $\beta$, which lead to reasonable model properties, have been found and fixed, we varied $A$ for choosing the ionization energy of the molecule. The potential has $N$ oscillations in the azimuthal direction, each minimum representing the location of one of the $N$ ions. The potential goes to zero for $\rho \rightarrow \infty$. 

\noindent It has been tested that the generation of harmonics is very weakly dependent on the fine details of the atomic potential. Instead, it is strongly dependent on its geometry and symmetries. It is therefore worth to look for a model potential that, keeping the proper symmetry, can be quite easily numerically optimized, i.e., requiring as few as possible grid points. In order to achieve this minimization of the number of grid points and to not break any physical symmetry we used a polar grid.  Moreover, we paid attention to use always a number of points in the azimuthal direction that is an integer multiple of $N$. We will take $N=6$, like in benzene,  and therefore the potential will exhibit a $C_6$ symmetry around the orthogonal axis.

\noindent For a good understanding of the harmonic spectra it is essential to study the level scheme in detail. Therefore, in order to characterize our model we have calculated the energy of the first six states for different potential strengths $A$ in the interval between 0.2 and 1.6. In this interval the energy of the ground state of the molecule decreases from $-0.14$ down to $-2.44$. The first, the third and the fourth states are non-degenerate, the others have a double degeneracy. In Fig.\,1 the energetic behavior of those states versus $A$ is shown. Contour plots of the six states for an intermediate value of $A = 0.80$ are shown in Fig.\,2. The pattern shape and the symmetries do not change with $A$, but the ``average radius'', i.e., the spatial extension, does. Clearly, for low values of $A$ the probability density is more loosely bound than for high $A$. In particular this is true for the upper states. For the sake of easy reference later on, we name the six states ${\bf 0}$, ${\bf 1}_a$, ${\bf 1}_b$, ${\bf 2}_a$, ${\bf 2}_b$, ${\bf 3}$, ${\bf 4}$, ${\bf 5}_a$ and  ${\bf 5}_b$. The subscripts are used just to distinguish among the degenerate states.  The non-degenerate states  ${\bf 0}$,  ${\bf 3}$ and ${\bf 4}$ have the full $C_6$ symmetry.

\noindent As one can infer from Fig.\,1 the six states can be divided in two branches: the first one, containing the first four states,  ${\bf 0}- {\bf 3}$, decreases very fast as $A$ increases, and a second one, containing the two upper states, ${\bf 4}- {\bf 5}$, which decreases much more slowly and lies even in the continuum for $A< 0.4$. As a result, for an increasing $A$ an increasing  gap between the two branches appears. To obtain the ionization potential of real benzene ($-0.34$ a.u.) we have to choose $A = 0.375$. Surprisingly, at that position the level scheme of our simple model resembles the molecular orbital (MO) scheme of the real benzene very well \cite{river}. In particular, the states ${\bf 4}$ and ${\bf 5}$ are still in the continuum so that only the four states ${\bf 0}- {\bf 3}$, possessing the same degeneracies as the MOs, are bound. A magnification of the region around $A = 0.375$ is also shown in Fig.\,1. Another parameter that will play a role in the HG is the level spacing $\Omega$  between the ground state and the first excited state. For an increasing $A$, i.e., a decreasing ground state energy, $\Omega$ decreases.

\section{Molecule-Field Interaction}

\noindent In dipole approximation the time-dependent Schr\"odinger equation for a single electron in a laser field $\vec{E}(t)$ and under the influence of an effective ionic potential $V(\vec{r})$ is given in length gauge by

\beq
\label{main}
i\frac{\partial}{\partial t}\Psi(\vec{r}, t) = \left ( -\frac{1}{2}\vec{\nabla}^2 + V(\vec{r}) +\vec{E}(t)\cdot \vec{r} \right ) \Psi(\vec{r}, t).
\eeq

\noindent In our case the dipole approximation is excellent because the molecule has a size much smaller than the wavelength of the laser field. We used a circularly polarized laser field that in cartesian coordinates is described by

\beq
\label{field}
\vec{E}(t) = \frac{{\cal{E}}(t)}{\sqrt{2}} \Big ( \mbox{cos}(\omega t)\,\vec{e}_x + \mbox{sin}(\omega t)\,\vec{e}_y \Big ),
\eeq

\noindent where ${\cal{E}}(t)$ is a slowly varying envelope with amplitude $\hat{{\cal E}}$, and $\omega$ is the laser frequency. In polar coordinates we obtain the TDSE 

\beq
\label{main2}
i\frac{\partial}{\partial t} \Psi(\rho, \varphi, t) = \left ( -\frac{1}{2\rho}\frac{\partial}{\partial \rho}\left( \rho\frac{\partial}{\partial \rho} \right ) - \frac{1}{2\rho^2}\frac{\partial^2}{\partial \varphi^2} + V(\rho, \varphi) + \frac{{\cal E}(t)}{\sqrt{2}} \rho\,\mbox{cos}(\varphi -\omega t) \right )\Psi(\rho, \varphi, t).
\eeq

\noindent This TDSE can be solved {\em ab initio} on a PC. We did this by propagating the wavefunction in time with a Crank-Nicholson approximant to the propagator $U(t + \Delta t, t) = \mbox{exp}[-i\Delta \, t H(t + \Delta t/2)]$ where $H(t)$ is the explicitly time-dependent Hamiltonian corresponding to the TDSE (\ref{main2}). Our algorithm is fourth order in the grid spacings $\Delta \rho$, $\Delta \varphi$ and second order in the time step $\Delta t$. The boundary condition is $\Psi(\rho,0,t) = \Psi(\rho,2\pi, t)$ for all $\rho$ and $t$. Probability density which approaches the grid boundary in $\rho$-direction is removed by an imaginary potential.

\section{Results} 
\noindent Here we discuss the results obtained from our 2D simulations and we compare them with  previous results from a one-dimensional model (1D) presented elsewhere \cite{lpb}. Our studies were mainly focused on the structure of the harmonic spectrum for different values of the ionization energy. In general, our findings show that, together with the harmonics we expected from the selection rules, other lines are present and their location can be, in most of the cases, explained with the help of the level scheme. In particular, we observed that for higher $A$ the gap between the two branches of Fig.\,1 plays an important role. For each $A$ various simulations with pulses of the same frequency and length but different intensities were performed. We used sine-square pulses of 30 cycles duration and a frequency $\omega = 0.0942$, unless noted otherwise. In order to better understand the additional lines which appear besides the expected harmonics it is useful to study the low intensity regime first.

\subsection{Low Fields}
\noindent In the case of a single atom, when the intensity of the field is not high enough for generating harmonics  efficiently, the Fourier transform of the dipole shows only the fundamental. Clearly, the threshold of the field strength to observe any harmonics depends on the ionization potential. In Fig.\,3 two dipole spectra, for different $A$, obtained from the interaction of the ring molecule with a low field pulse are shown. For the dipole emission spectrum of Fig.\,3a an electric field amplitude $\hat{\cal E} = 0.005$ was used and $A=1.6$. In this case apart from the fundamental and the weak fifth harmonic, four other lines are present. Two lines are located at $14.1\, \omega$ and $16.0\, \omega$, respectively. We refer to them as $\Lambda_a$ and $\Lambda_b$ hereafter. These two lines correspond to two resonances between the states ${\bf 0} \rightarrow {\bf 5}$ and  ${\bf 3} \rightarrow {\bf 4}$ and therefore they move towards the red if $A$ decreases. This is confirmed in Fig.\,3b, where $A = 1.4$ and $\hat{\cal E} = 0.005$. In Fig.\,3b the 5th harmonic is more pronounced. This is due to the fact that when the ionization potential is lower the generation of harmonics requires weaker fields. A subfundamental line at $\Omega$, corresponding to a transition from the first excited state ${\bf 1}$ to the ground state is also present in both cases. We name this  line $\Upsilon$. Another line corresponding to the transition $2\omega-\Omega$ is visible as a right satellite to the fundamental. Looking closer, one observes that, with respect to the positions one would expect from the unperturbed level scheme, $\Lambda_a$ and $\Lambda_b$ are blue-shifted, whereas $\Upsilon$ is not. This opposite shift can be explained by the dynamical Stark effect. The ground state remains almost unaffected and the lower-lying states move slightly. The higher-lying states instead, experience a relatively strong shift. Therefore, with increasing laser intensity the gap between the two branches in Fig.\,1 increases, leading to a blue shift of $\Lambda_a$ and $\Lambda_b$. The three lines $\Lambda_a$, $\Lambda_b$ and $\Upsilon$ are the first lines to appear in the low field regime. The subfundamental line can be considered as characteristic for the low field regime. Indeed, with increasing field strength, it moves towards the red and finally becomes very difficult to be resolved.

\subsection{High Fields}
\noindent With increasing laser intensity the actual harmonic spectrum develops. In Fig.\,4 we show HG spectra for four different electric field amplitudes and $A = 1.6$. In Fig.\,4a $\hat{\cal E} = 0.02$ a.u. and the situation is, at first sight, quite similar to that one of Fig.\,3a: $\Lambda_a$ and $\Lambda_b$ are present, and the first allowed harmonic is there, i.e. the 5th. Increasing the field to $\hat{\cal E} = 0.12$ a.u., Fig.\,4b, also the $7$th and the $11$th appear, and the two lines $\Lambda_a$ and $\Lambda_b$ are still present but located within a broader structure. This last phenomenon can be explained taking into account at least three different effects: (i) in general, the width of $\Lambda_a$ and $\Lambda_b$ increases for higher fields, (ii) other channels, i.e. other resonance lines between the two branches, can be opened if the field becomes intense, and (iii) through a removal of the degeneracy of the states by the electric field more possible resonance lines are obtained. These factors generate a kind of broad ``hill'' in the harmonic spectrum, the position of which is a function of $A$. Moreover, in Fig.\,4b other satellite lines around the expected harmonics are present. Those lines are decays from virtual states to real states. In Fig.\,4c, $\hat{\cal E}=0.21$ a.u., more couples of harmonics are present and the hill is reduced to a light background modulation of the main HG spectrum. This becomes even clearer in Fig.\,4d where harmonics up to the $49$th are observed and the $17$th is located just on the hill. We have also studied in more detail how the strength of the harmonics increases (or decreases) in function of the electric field. The results for harmonics up to the 31st are shown in Fig.\,5. For each couple of harmonics there is a minimum field threshold, below which the harmonic lines  cannot be picked out from the background. All the harmonic strengths of Fig.\,5 are normalized to the fundamental. What is worth to stress is that for the 5th harmonic we can have an efficiency up to 13\%. 

\noindent Repeating the sequence of simulations shown in Fig.\,4 for a lower $A$ gives results that are quite similar to those shown in Fig.\,4 as long as the gap between the two branches is quite large. When the gap becomes of the order of about ten photons the two lines $\Lambda_a$ and $\Lambda_b$ play a role that is less important. This is due to the fact that, as the gap is smaller, the two lines are expected to be located in a low frequency region and therefore they are easily hidden by the background of the main HG spectrum that in the low frequency region is higher. Furthermore, as the intensity of the harmonic lines is strongly enhanced with increasing field, the strength of $\Lambda_a$ and $\Lambda_b$ is not. Also the extension of the harmonic spectrum is dependent on $A$, for lower $A$ less harmonics can be generated (for a fixed frequency).

\noindent As already mentioned, $\Omega$ is the distance between ${\bf 0}$ and ${\bf 1}$. For high $A$ (or high $\omega$) we have $\omega > \Omega$ but for decreasing $A$ (or decreasing $\omega$), $\Omega$  approaches the laser frequency and overtakes it. The ratio of $\omega$ and $\Omega$ strongly affects the harmonic emission by the ring molecule. In particular, we observed that when $\omega \approx \Omega$ a very complex spectrum is generated and together with the expected harmonics many other lines of similar intensities are present. This effect can be seen in Fig.\,6. A very similar behavior was also observed in the 1D model \cite{lpb}. In particular, the shape of the additional satellite structures around each allowed harmonic are in the two cases alike. It seems that in this resonant case the system is not in a single non-degenerate Floquet state as assumed for the proof of the selection rules \cite{alon}. In that derivation the pulse was assumed as infinite. Therefore, the particular behavior of the dipole emission could be also due to a pulse shape effect. However, pulse shape effects should be not dependent on the frequency, and in fact, we have additional lines for all the frequencies, but those lines play always a minor role with respect to the expected harmonics. Instead, when the laser frequency becomes nearly resonant many new strong lines appear.

\noindent Keeping the same pulse parameters and decreasing the parameter $A$ the ionization increases. When we want to study a model that is closer to the real benzene, we have to take $A = 0.375$. With this condition the physical scheme is very different from those of the cases previously discussed; the ionization energy is reduced from $2.44$ to $0.34$ and, as we already mentioned, only four states are present and the gap between level ${\bf 3}$ and ${\bf 4}$ does not exist at all. Under these conditions a pulse with the frequency $\omega=0.0942$, which we used so far, leads to emission spectra without any harmonic structure. This is mainly for the reason that the frequency is relatively high with respect to the ionization potential and therefore just a very few photons are sufficient for reaching the continuum. Therefore, unless the field is very low the ionization would prevail at soon. Also making a comparison with the rule for the cut-off position used in the atomic case ($I_p + 3.17 U_p$) we should not expect any harmonics due to the low value of $U_p$. Therefore, we made a series of simulations with the benzene model but using a lower frequency, $\omega = 0.0314$. With this frequency eleven  photons are required for reaching the continuum, i.e., the molecule can be ionized only with a high-order multiphoton process. A spectrum obtained with this low frequency and $\hat{\cal E} = 0.044$ is shown in Fig.\,7. Like in the highly-bound high frequency case the emission spectrum exhibits the harmonics allowed by the selection rules. The efficiency of the harmonics in Fig.\,7 is not as high as the one of the harmonics of Fig.\,4. Another line, located around $3\,\omega$  is also present in the spectrum of Fig.\,7. This line is a resonance between the states ${\bf 0}$ and ${\bf 1}$ (for $A = 0.37$, $\Omega = 0.0914 \approx 3\,\omega$ ). It is interesting to note that in this case of a weakly bound electron the results from the 2D simulations are different with respect to those from the 1D simulations \cite{lpb}. In the latter case no harmonic structure was observed for $\omega < \Omega$. This, in our opinion, could be due to the reason that in the 1D model the level scheme is qualitatively different. In particular, there is no continuum in the 1D case. 

\section{Conclusions}

\noindent In this work we have studied the harmonic emission in a ring molecule. We have shown that when a ring molecule interacts with a laser pulse, together with the series of harmonics predicted by the selection rules, other lines are present. Under certain conditions the strength of these lines can be comparable with that one of the harmonics. Our HG spectra present a structure and a complexity that is absent in the numerical results shown in \cite{alon}. This is due to the reason that while there a 1D Floquet simulation was performed in our studies a realistic pulse (i.e., a finite pulse with a certain envelope) and a 2D model with ionization included were used, and the TDSE was solved {\em ab initio}. 

\noindent What is worth to note is the scaling of the TDSE (\ref{main2}) with respect to the size of the molecule. If one scales the molecule radius like $\rho ' = \alpha \rho$, the TDSE (\ref{main2}) remains invariant if $t' = \alpha^2 t$, $V' = V/\alpha^2$, $\hat{\cal E}' = \hat{\cal E}/\alpha^3$, $\omega' = \omega/\alpha^2$ are chosen. Therefore our results for high $A$ can reproduce the results that would have been obtained for a bigger molecule with a lower ionization potential interacting with a field of lower frequency. Moreover, one can think about the different cases of Fig.\,1 as the level scheme of positively charged molecules as well. 

\noindent To our knowledge, so far harmonics in ring-shaped molecules have been investigated experimentally only in gaseous samples \cite{organic}. Studying a gaseous sample is very different with respect to what we did in our simulations. Because the molecules in a gas have a random orientation it is not possible to apply the symmetry properties discussed. If the propagation direction of the circularly polarized field is not orthogonal to the plane of the molecule, the molecule ``sees'' a field that is elliptically polarized. This breaks the discrete rotational symmetry and other harmonics become possible. Nonetheless, even if it is not possible to make a direct comparison it is useful to note that the results presented in \cite{organic} confirm that ring  molecules, like benzene and cyclohexane, can tolerate short pulses of high intensity and phenomena like fragmentation  and Coulomb explosion do not play a big role. In order to reproduce in a real experiment the results we presented, it is fundamental to prepare a sample where most of the molecules lie in the same plane. This could be done with some orientation techniques or, considering the particular shape of the organic molecules, preparing a very thin layer.

\noindent We described and discussed results for a molecules with $N = 6$, but cases with higher $N$ are as well possible. Moreover, the higher is $N$ the higher is the frequency of the first generated harmonic; in the limit of very high $N$, harmonics of very short wavelength could be generated. 

\noindent In this work we a took into account a single active electron, but also including correlation through both, a full description \cite{alon} or an approximate treatment like time-dependent density functional theory, would not change the symmetry properties of the Hamiltonian describing the system. Therefore, the selection rules would apply as well.

\noindent More complex molecules which produces similar selection rules \cite{alon2} are the nanotubes \cite{saito}. They can be very long in the longitudinal direction and exhibit a discretized cylindrical symmetry. A semiclassical approach to harmonic generation from nanotubes was also investigated \cite{slepyan}. Unfortunately, because of their size, the dipole approximation would not be accurate enough. Therefore, an {\em ab initio} numerical simulation in 3D  would be at the limit, or probably beyond, the calculation capabilities of even the fastest computers now available. 
 
\begin{center} {\bf ACKNOWLEDGEMENTS} \end{center}

\noindent This work was supported in part by the Deutsche Forschungsgemeinschaft within the SPP ``Wechselwirkung intensiver Laserfelder mit Materie''.

\newpage
\begin{center} {\Large FIGURES} \end{center}

\noindent FIG. 1. Energetic behavior of the first six states versus $A$. For lower $A$ some of the states belong to the continuum; the magnification shows the region around $A = 0.375$ where four states are bound. For higher $A$ the set of states is split in two branches and a gap between those branches appears.\\

\noindent Fig. 2. Contour plots of the first six states. For each double degenerate state, two linearly independent states are shown. The non-degenerate states present a full $C_6$ symmetry.\\

\noindent Fig. 3. Emission spectrum in the low field regime. In (a) $A = 1.6$ and $\hat{\cal E} = 0.005 $, in (b) $A = 1.4$ and $\hat{\cal E} = 0.005$. The lines $\Lambda_a$, $\Lambda_b$ and $\Upsilon$ are present in both pictures. When $A$ decreases the two lines $\Lambda_a$ and $\Lambda_b$ are red-shifted. This can be observed comparing (b) with (a).\\

\noindent Fig. 4. Evolution of the harmonic spectrum with increasing field for $A = 1.6$. In (a) $\hat{\cal E} = 0.02$, in (b) $\hat{\cal E} = 0.12$, in (c) $\hat{\cal E} = 0.21$ and in (d) ${\cal E} = 0.27$. When the field increases additional satellite lines appear with the low order harmonics. The highest allowed resolved harmonic is the 49th.\\

\noindent Fig. 5. Strength of each harmonic line versus the electric field ${\cal E}$. It is worth to note that the first allowed harmonic, i.e., the 5th, can reach an efficiency up to 13\%. In the low field region the strength of the the 13th and 17th harmonic is particularly high, this is due to the presence of the ``hill''. The value of $A$ is constant, $A = 1.6$. \\

\noindent Fig. 6. Emission spectrum for $A = 1.00$ and $\omega = 0.0942$. The value of $\Omega$ approaches the laser frequency $\omega$ and the spectrum exhibits a complex structure with many additional strong lines. \\

\noindent Fig. 7. Emission spectrum for $A = 0.375$ and $\omega = 0.0314$. Also in this case, where the ionization potential corresponds to that one of the benzene molecule, the spectrum exhibits the same structure as in the highly-bound high frequency case. The line around $3\,\omega$ is given by a decay from the state ${\bf 1}$ to the state ${\bf 0}$.

\newpage
\pagestyle{empty}
\setlength{\unitlength}{1.0cm}
\begin{picture}(9.0,4.0)
\put(-3.5,-23.5){\bild{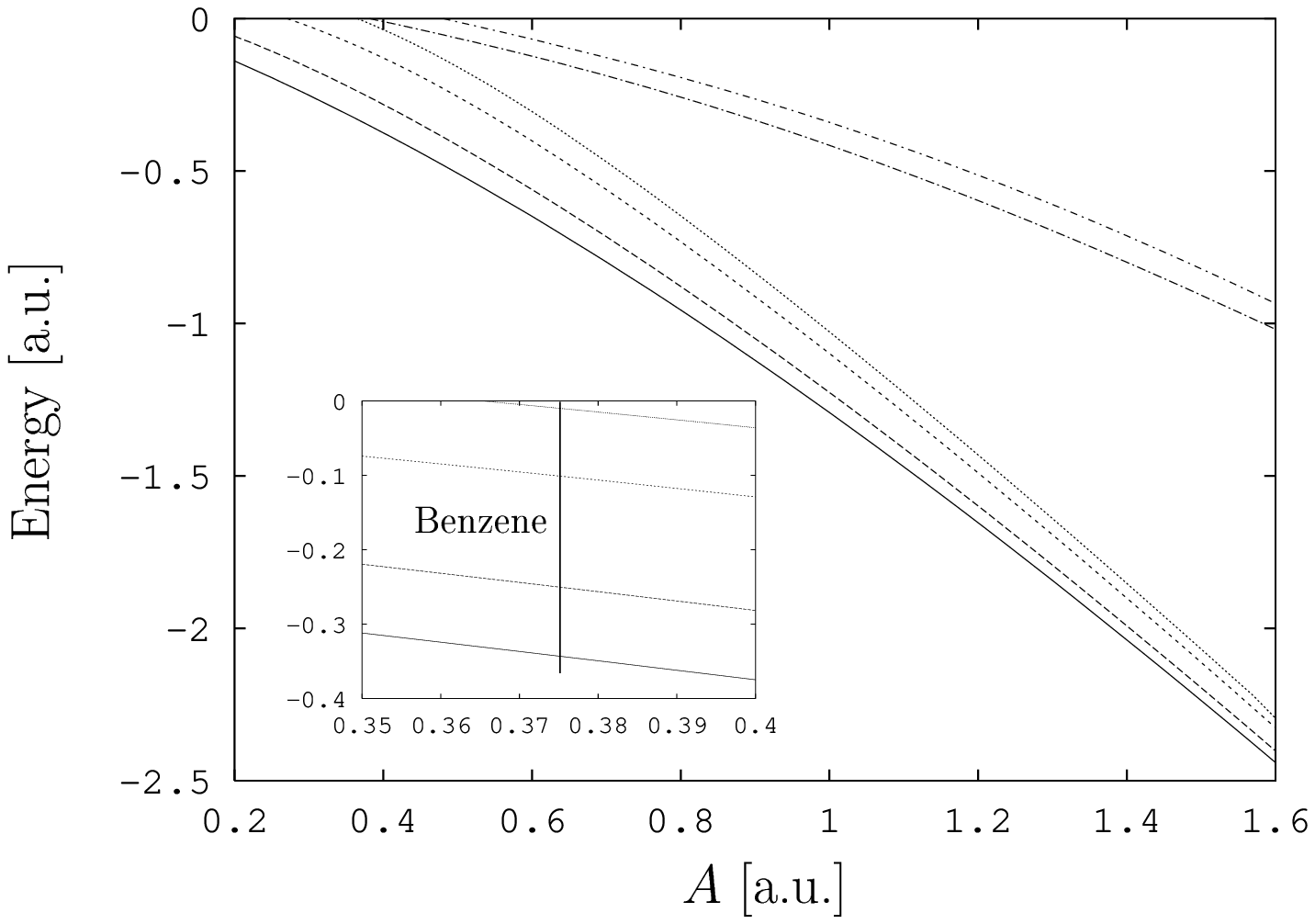}{22.0cm}}
\end{picture}

\vspace*{18cm}
\noindent {\bf Fig. 1: F.\,Ceccherini and D.\,Bauer, ``Harmonic generation in ...''}

\newpage
\pagestyle{empty}
\setlength{\unitlength}{1.0cm}
\begin{picture}(9.0,4.0)
\put(0.0,-9.5){\bild{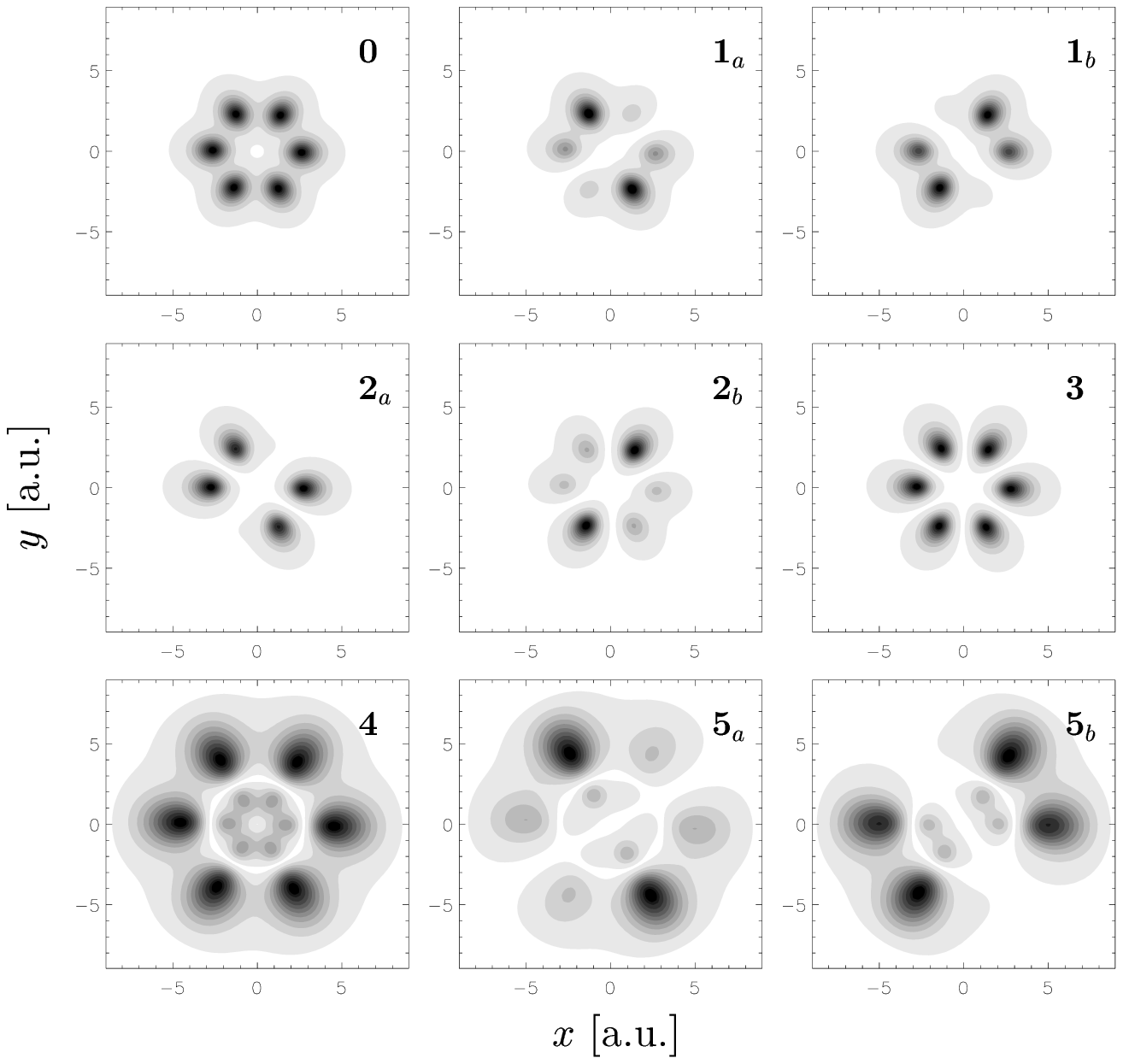}{16.0cm}}
\end{picture}

\vspace*{18cm}
\noindent {\bf Fig. 2: F.\,Ceccherini and D.\,Bauer, ``Harmonic generation in ...''}

\newpage
\begin{picture}(9.0,4.0)
\put(-3.0,-21.5){\bild{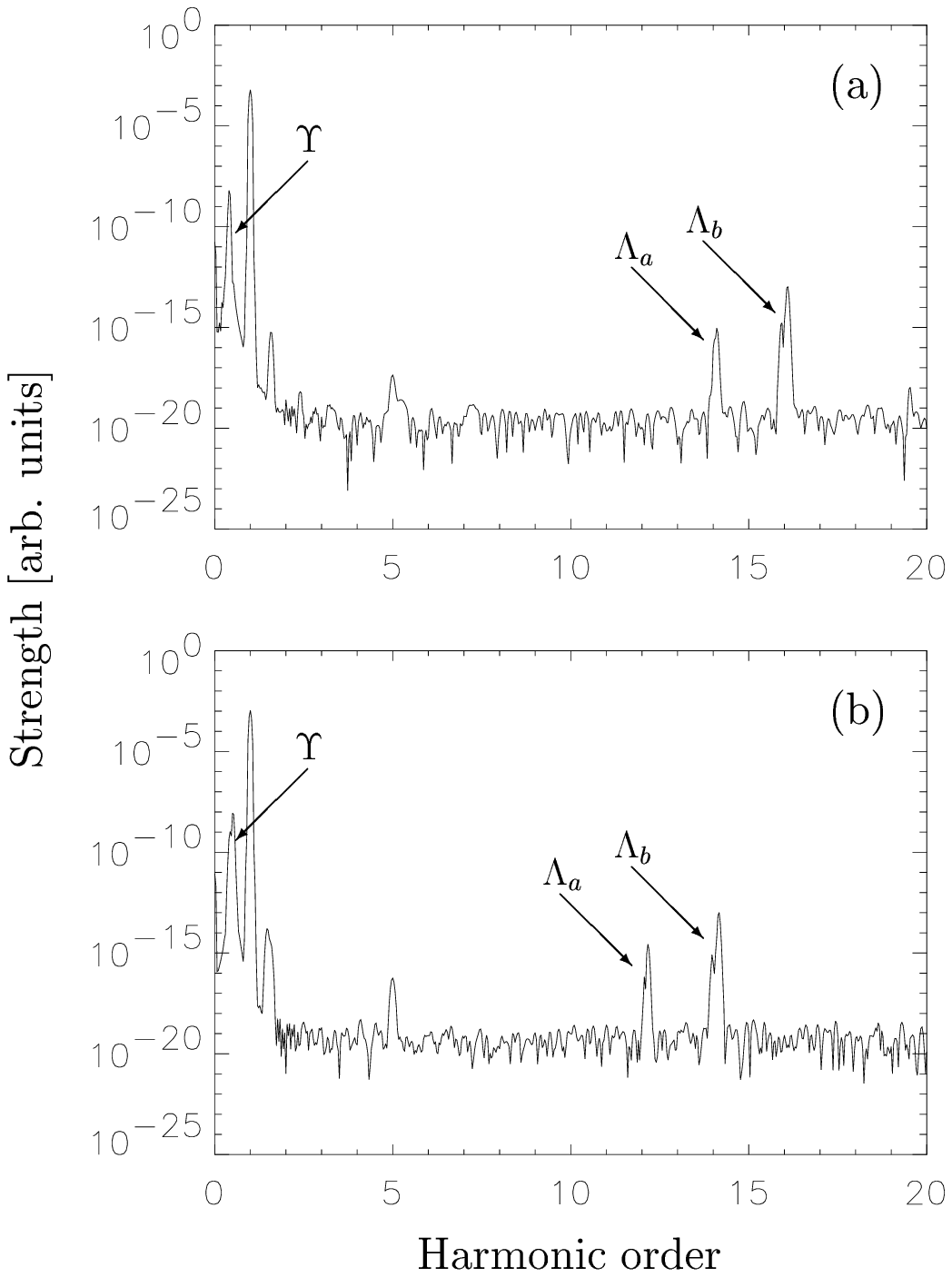}{20.0cm}} 
\end{picture}

\vspace*{18cm}
\noindent {\bf Fig. 3: F.\,Ceccherini and D.\,Bauer, ``Harmonic generation in ...''}

\newpage
\begin{picture}(9.0,4.0)
\put(-3.0,-23.8){\bild{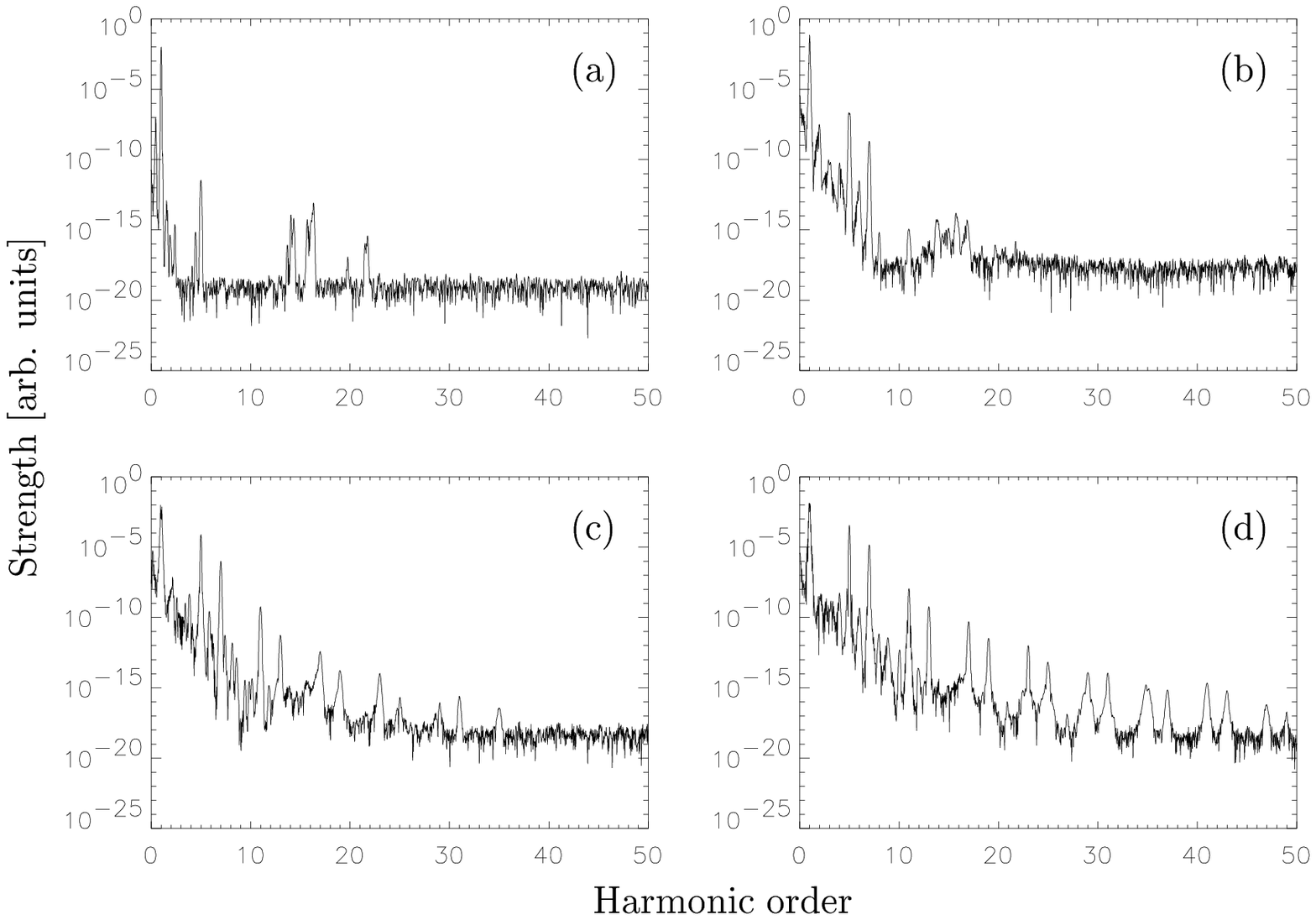}{22.0cm}} 
\end{picture}

\vspace*{18cm}
\noindent {\bf Fig. 4: F.\,Ceccherini and D.\,Bauer, ``Harmonic generation in ...''}

\newpage
\pagestyle{empty}
\setlength{\unitlength}{1.0cm}
\begin{picture}(9.0,4.0)
\put(-2.0,-20.0){\bild{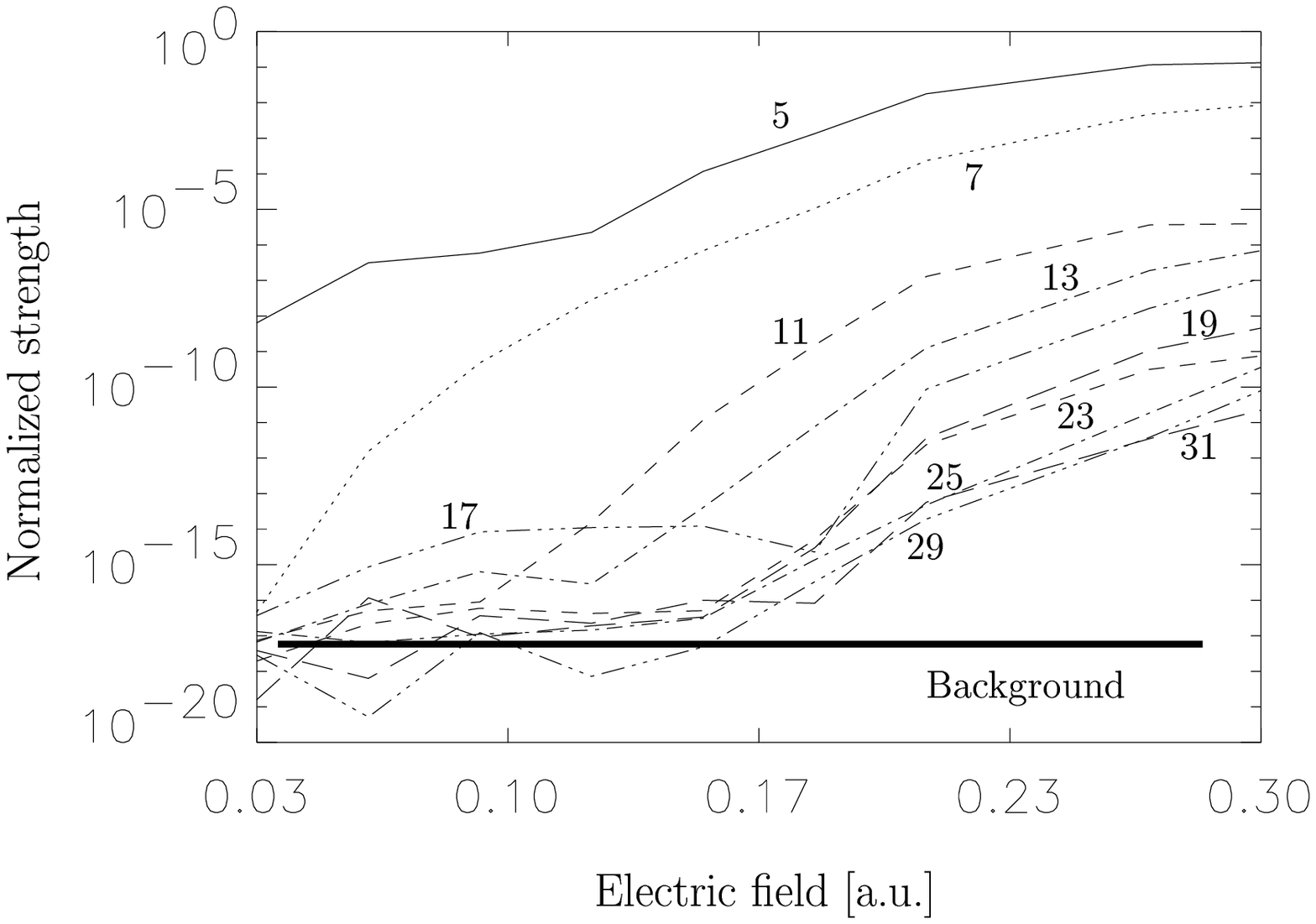}{19.0cm}}
\end{picture}

\vspace*{18cm}
\noindent {\bf Fig. 5: F.\,Ceccherini and D.\,Bauer, ``Harmonic generation in ...''}

\newpage
\begin{picture}(9.0,4.0)
\put(-2.0,-20.0){\bild{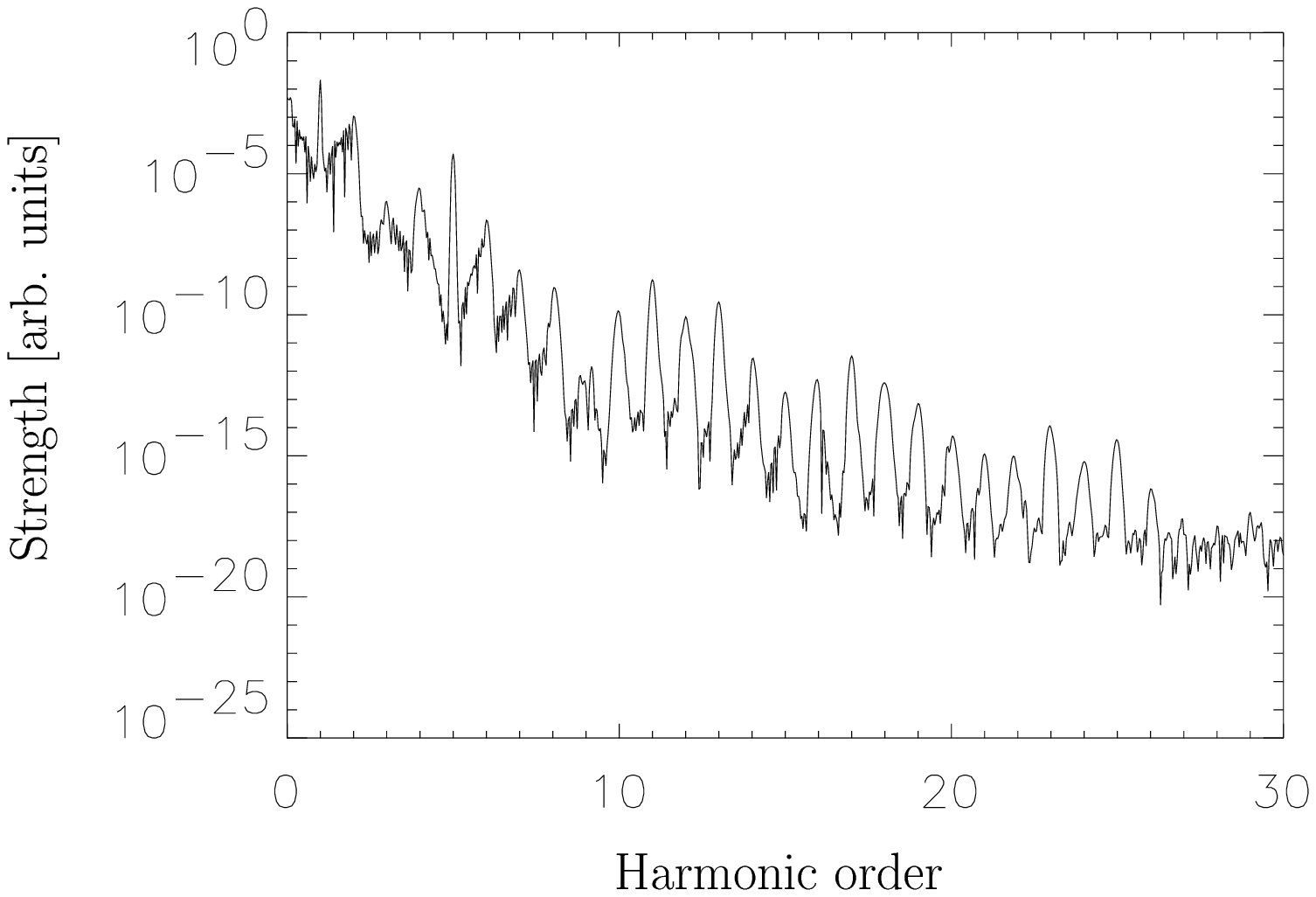}{19cm}}
\end{picture}

\vspace*{18cm}
\noindent {\bf Fig. 6: F.\,Ceccherini and D.\,Bauer, ``Harmonic generation in ...''}

\newpage
\begin{picture}(9.0,4.0)
\put(-2.0,-20.0){\bild{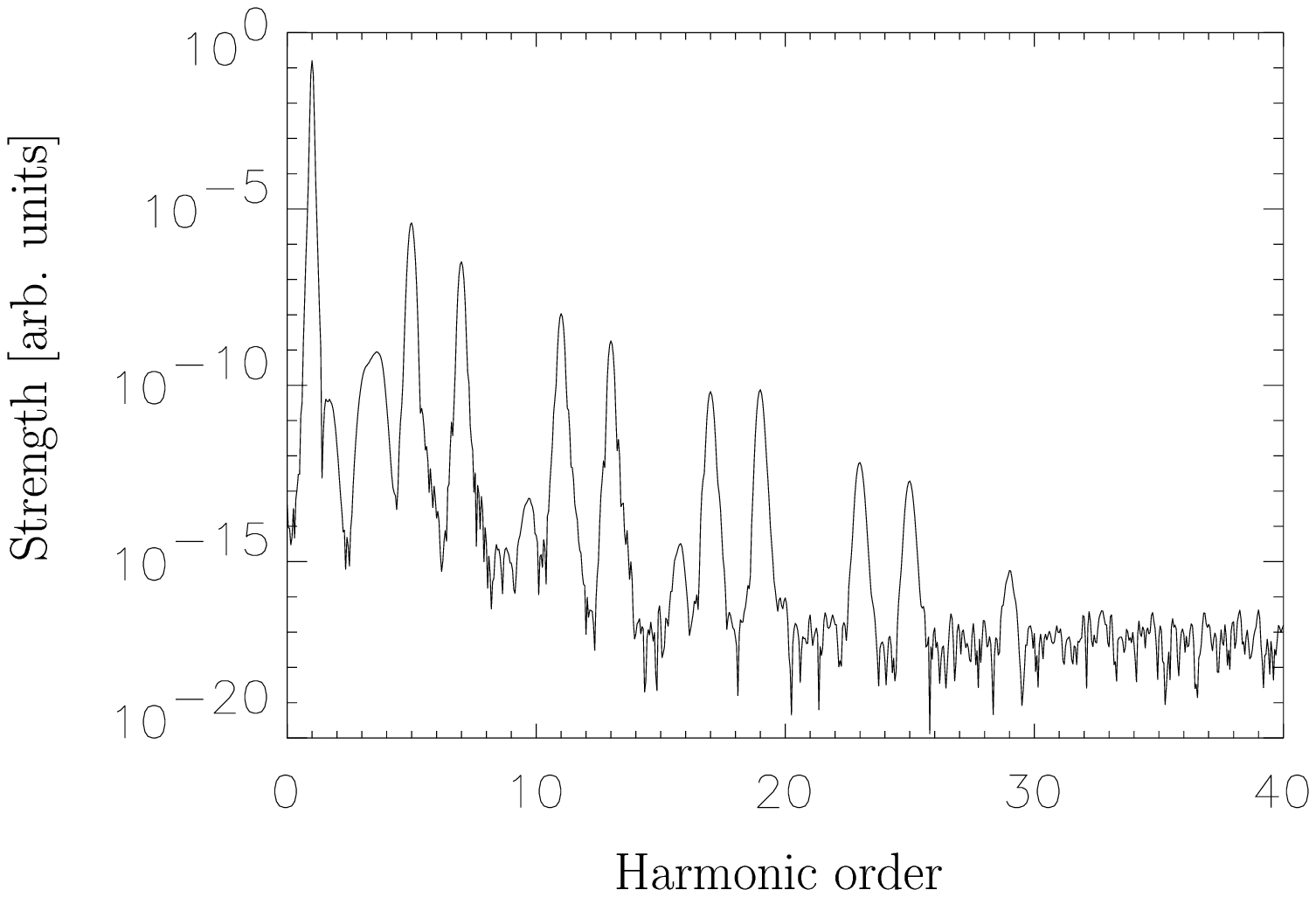}{19cm}}
\end{picture}

\vspace*{18cm}
\noindent {\bf Fig. 7: F.\,Ceccherini and D.\,Bauer, ``Harmonic generation in ...''}


\begin{thebibliography}{99}
\bibitem{smooth} The need of a ``smoothed'' potential is due to the singularities that otherwise in the 2D simulation would be present at the location of each atom.
\end{thebibliography}
\end{document}